\documentclass{article}
\usepackage{srcltx}
\usepackage{amsmath}
\usepackage{amssymb}
\usepackage{amsfonts}
\usepackage{latexsym}
\usepackage{textcomp}
\usepackage{multirow}
\usepackage{booktabs}
\usepackage{url}
\usepackage{array}
\usepackage{marvosym}
\usepackage{graphics}
\usepackage{graphicx}
\usepackage[caption=false]{subfig}
\usepackage{adjustbox}
\usepackage{epsfig}
\usepackage[round,authoryear]{natbib}

\usepackage{lineno}

\graphicspath{{./figures/}}

\title{The inefficiency of Bitcoin revisited: a dynamic approach}

\author{Aurelio F. Bariviera \\ 
{\small Department of Business, Universitat Rovira i Virgili, Av. Universitat 1, 43204 Reus, Spain}\\
{\small Escuela de Postgrado, Universidad del Pac\'ifico, Av. Salaverry 2020, Lima, Per\'u}\\
{\small aurelio.fernandez@urv.cat}
}

\begin{document}

\maketitle

\begin{abstract}
This letter revisits the informational efficiency of the Bitcoin market. In particular we analyze the time-varying behavior of long memory of returns on Bitcoin and volatility 2011 until 2017, using the Hurst exponent. Our results are twofold. First, R/S method is prone to detect long memory, whereas DFA method can discriminate more precisely variations in informational efficiency across time. Second, daily returns exhibit persistent behavior in the first half of the period under study, whereas its behavior is more informational efficient since 2014. Finally, price volatility, measured as the logarithmic difference between intraday high and low prices exhibits long memory during all the period. This reflects a different underlying dynamic process generating the prices and volatility.\\
\textbf{Keywords:} Bitcoin; Long range dependence; volatility; Hurst exponent\\
\textbf{JEL Classification:} G01; G14
\end{abstract}

\section{Introduction}
In recent years, a new type of financial asset, was introduced. This new type is labeled with the generic name of cryptocurrency.  The most popular one is Bitcoin, developed upon the seminal paper by \cite{Nakamoto}. Its market capitalization is 15 USD billions, it is traded with many of the main national currencies, with a daily turnover of more than 17 USD milions \citep{coinmarketcap}, highlighting the importance of this market from an economic perspective.
 
Bitcoin emerged recently as a new topic in empirical economic studies. Scopus database (as of September 2017) includes 742 documents with 'bitcoin' in its title or keywords. These articles cover legal, economic or computer aspects of the cryptocurrency. For brevity, we will make a brief review of recent articles on economic and statistical aspects of bitcoin. \cite{DonierBouchaudPlosOne} study different measures of liquidity as early warning signs of bitcoin market crash. Two recent papers in this journal examined the informational efficiency of the Bitcoin market. In particular, \cite{Urquhart2016} used a set of tests aimed at identifying autocorrelations, unit roots, nonlinearities and long range dependence in Bitcoin returns. The results are compelling for the whole period under study, rejecting the null hypothesis of a random behavior of the time series. However, when splitting the examination into two non overlapping periods, the paper uncovers that the first subperiod is the main responsible for the informational inefficiency. Later, \cite{Nadarajah2017} reexamines the data using power transformations of daily returns, without rejecting the null hypothesis of informational efficiency. \cite{Bourietal2017} study bitcoin's return-volatility behavior before and after the severe market crash of 2013., and find serial correlation in bitcoin return series. \cite{Bouri20171frl} scrutinize hedge and safe haven properties of Bitcoin \textit{vis-\`a-vis} several stock, bonds and currency indices around the world. Its main finding is that the cryptocurrency is only useful as a diversifier device, but not as a hedge instrument. Finally, \cite{Balcilar20177em} detect nonlinearities in the return-volume relationship, which allows for return prediction.

This letter improves and complements previous literature on Bitcoin in three aspects: (i) it proposes the use of Detrended Fluctuation Analysis method, instead of the commonly used R/S method, (i) it works with sliding windows, in order to dynamically assess the evolution of informational efficiency across time and (ii)  it considers, in addition to daily returns, the long range memory in the daily volatility of returns, which can proxy the risk of this unstable market. Volatility analysis is particularly important, because, as studied by \cite{Blau2017} speculative trading was not responsible for the high volatility of Bitcoin during the period 2010-2014.
 
The letter is organized as follows. Section \ref{sec:EMH} briefly review some key aspects of the Efficient Market Hypothesis. Section \ref{sec:data} presents the data and methodology. Section \ref{sec:results} discusses the main findings. Finally, Section  \ref{sec:conclusions} outlines the conclusion of our analysis.

\section{The Efficient Market Hypothesis \label{sec:EMH}}
Over a century ago \citet{Bachel} developed the first mathematical model of security prices, applying the arithmetic Brownian motion model to French bonds. The formalization of the Efficient Market Hypothesis (EMH) remained latent until its theoretical development by \citet{Samuelson65} and the definition and classification by \citet{Fama70}. Briefly, the EMH requires that returns of financial assets follow a memoryless stochastic process with respect to the underlying information set. 

The weak form of informational efficiency excludes the possibility of finding, systematically, profitable trading strategies. As a corollary, returns time series cannot exhibit predictable memory content. However, there are several studies that find long memory in financial time series, using different methods. For example, \citet{BarkoulasBaumTravlos00} and \citet{BlascoSantamaria96} find long memory in the Greek Stock Exchange and Spanish Stock Exchange respectively. \citet{CheungLai95} find empirical evidence of long memory in 5 out of 18 developed stock markets and \citet{BarkoulasBaum96} do not find strong convincing evidence against the random walk model in US stock returns.
In spite of the fact that fixed income instruments are very important in the composition of investment portfolios and in firm and government financing, they have been less studied than stocks. \citet{Carbone04} find long memory in German stock and sovereign bond markets and \citet{McCarthy09} find long memory in yields of corporate bonds and in the spread of returns of corporate bonds and treasury bonds.

Another issue in the literature is the time varying behavior of the market efficiency. The reasons for the varying memory remains a puzzle. In this aspect \citet{ItoSugiyama09} find that inefficiency varies through time in the US stock market. \citet{Bariviera11} finds that time varying long-range dependence in the Thai Stock Market is weakly influenced by the liquidity level and market size. \citet{CajueiroGogasTabak09} find that financial market liberalization increases the informational efficiency in the Greek Stock Market. \citet{KimShamsuddinLim11} find that return predictability is altered by political and economic crises but not during market crashes. Remarkably, there is no paper dealing with the long memory of Bitcoin return volatility.

\section{Data and methodology \label{sec:data}}
We use daily price data of Bitcoin. All data used in this paper was retrieved from DataStream. The period under examination goes from 18/08/2011 to 15/02/2017, for a total of 1435 observations. 

The Hurst exponent $H$ characterizes the scaling behavior of the range of cumulative departures of a time series from its mean. The study  of long range dependence can be traced back to seminal paper by \citet{Hurst51}, whose original  methodology was applied to detect long memory in hydrologic time series. This method was also explored by \cite{Mandel68} and later introduced in the study of economic time series by \cite{Mandel72}.  This method uses the range of the partial sums of deviations of a time series from its mean, rescaled by its standard deviation. If we have a sequence of continuous compounded returns $\{r_1, r_2, \dots , r_{\tau} \}$, $\tau$ is the length of the estimation period and $\bar{r}_{\tau}$ is the sample mean, the the $R/S$ statistic is given by
\begin{equation}
(R/S)_\tau \equiv \frac{1}{s_\tau} \left[ \max_{1\leq t\leq\tau}\sum_{t=1}^\tau (r_t-\bar{r}_{\tau})- \min_{1\leq t\leq\tau}\sum_{t=1}^\tau(r_t-\bar{r}_{\tau})\right]
\label{eq:RS}
\end{equation}
where $s_{\tau}$ is the standard deviation
\begin{equation}
s_\tau\equiv \left[\frac{1}{\tau}\sum_{t}(r_t-\bar{r}_\tau)^2\right]^{1/2}
\label{sdtRS}
\end{equation}
Hurst \cite{Hurst51}, found that the following relation
\begin{equation}
(R/S)_\tau =(\tau/2)^H
\label{eq:H}
\end{equation}
is verified by many time series in natural phenomena.

Since then, several methods (both parametric and non-parametric) have been developed to calculate the Hurst exponent.  We advocate for the use of the Detrended Fluctuation Analysis (DFA) method developed by \citet{Mosaic94} because, as highlighted by \citet{GrauCarles}, it avoids the spurious detection of long-range dependence. 
The algorithm, described in detail in \cite{Peng95}, begins by computing the mean of the stochastic time series $y(t)$, for $t=1,\dots, M$. Then, an integrated time series $x(i)$, $i=1,\dots, M$ is obtained by subtracting mean and adding up to the $i-th$ element, $x(i)=\sum_{t=1}^{i}[y(t)-\bar{y}]$. Then $x(i)$ is divided into $M/m$ non overlapping subsamples and a polynomial fit $x_{pol}(i,m)$ is computed in order to determine the local trend of each subsample. Next the fluctuation function 
\begin{equation}
F(m)=\sqrt{\frac{1}{M}\sum_{i=1}^{M}{[x(i)-x_{pol}(i,m)]}^2}
\label{eq:DFA}
\end{equation}
is computed. This procedure is repeated for several values of $m$. The fluctuation function $F(m)$ behaves as a power-law of $m$, $F(m) \propto m^H$, where $H$ is the Hurst exponent. Consequently, the exponent is computed by regressing $\ln(F(m))$ onto $\ln(m)$. According to the literature the maximum block size to use in partitioning the data is $(length(window)/2)$, where \textit{window} is the time series window vector. Consequently, in this paper we use six points to estimate the Hurst exponent. The points for regression estimation are: $m=\{4, 8, 16, 32, 64, 128\}$.  For a survey on the different methods for estimating long range dependences see \citet{Taqqu95}, \citet{Montanari99} and  \citet{Serinaldi10}. 

Departing from daily returns,  we estimate the Hurst exponent using 500 datapoints sliding window. This rolling sample approach works as follows: we compute the Hurst exponent for the first 500 returns, then we discard the first return and add the following return of the time series, and continue this way until the end of data. Thus, each $H$ estimate is calculated from data samples of the same size. We obtained an average of 935 Hurst estimates.  This sliding window methodology was successfully used to assess time varying informational efficiency in  \citet{CTchina,CajueiroGogasTabak09,CajueiroTabak2010,Bariviera11,BaGuMa12,BaGuMa14,Barivieraetal2017}, among others.

We compute the Hurst exponent by both R/S and DFA methods for the usual daily logarithmic return:
\begin{equation}
r_t=(\ln P_t-\ln P_{t-1})*100
\end{equation}
and for the daily price volatility, defined as the logarithmic difference between intraday highest and lowest prices:
\begin{equation}
ReturnVolatility_t=\ln P_t^{high}-\ln P_{t}^{low}
\end{equation}

\begin{table}[]
\centering
\caption{Descriptive statistics of bitcoin returns and volatility.}
\label{return-statistics}
\begin{tabular}{rrr}
\toprule
              & Return  & Volatility  \\
              \midrule
Observations &  1434   & 1434    \\
Min     &  -66.3900  &   -102.9000  \\
Max     &   48.4800   & 179.0000\\
Mean    &    0.3159   & 6.2810\\
Median   &    0.2310  & 3.8680\\
Std Deviation & 6.2104   & 10.1116 \\
Skewness &  -1.1833  &  6.5764 \\
Kurtosis  & 25.5773 &   107.5844 \\
Jarque Bera & 30791$^*$ & 663412  $^*$ \\
\bottomrule
\multicolumn{3}{l}{$^*$ Significant at 1\% level.}\\
\end{tabular}
\end{table}

Table  \ref{return-statistics} presents the descriptive statistics of the return and volatility time series. We observe, similarly to other financial assets, that these series are leptokurtic and non-normal. 

\section{Results \label{sec:results}}

\begin{figure}[h!p]
    \centering
        \subfloat[Hurst exponent of BTC daily returns.]{%
           \label{fig:hursreturn}
        \includegraphics[scale=0.65]{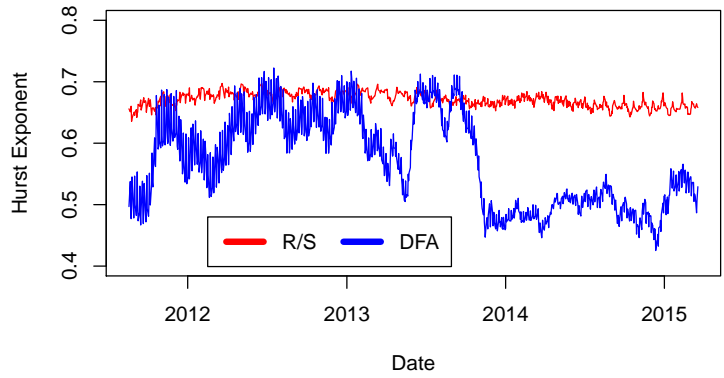}
            } \\%
        \subfloat[Hurst exponent of BTC daily return volatility.]{%
            \label{fig:hurstvolatility}
        \includegraphics[scale=0.65]{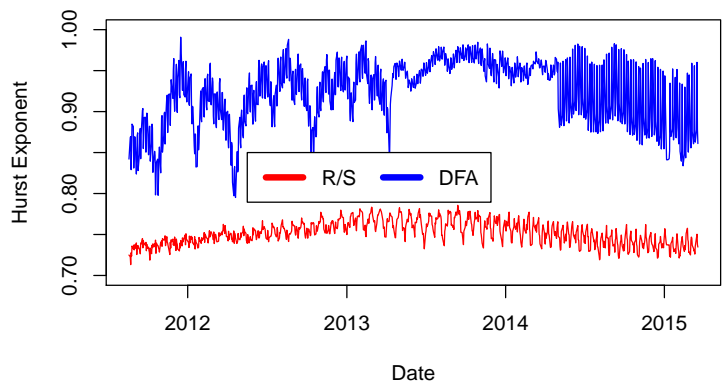}
        }
    \caption{Hurst exponent using DFA method (blue) and R/S method (red), using sliding windows of 500 datapoints and one datapoint step forward. The date corresponds to the first observation of the sliding window.}
   \label{fig:hurst}
\end{figure}

The use of the sliding window allows the observation of a time-varying pattern in long memory in financial time series. The graphical results of our empirical study is presented in Figure \ref{fig:hurst}. 

\begin{table}[]
\centering
\caption{Descriptive statistics of Hurst estimates.}
\label{Hurst-statistics}
\begin{tabular}{rrrrr}
\toprule
 & \multicolumn{2}{c}{Hurst R/S}    &        \multicolumn{2}{c}{ Hurst DFA}  \\
              & Return  & Volatility & Return  & Volatility \\
              \midrule
Observations  & 935     & 935        & 935     & 935        \\
Min           & 0.6357  & 0.7132     & 0.4253  & 0.7952     \\
Max           & 0.6974  & 0.7856     & 0.7224  & 0.9908     \\
Mean          & 0.6711  & 0.7514     & 0.5698  & 0.9221     \\
Median        & 0.6716  & 0.7509     & 0.5660  & 0.9321     \\
Std Deviation & 0.0116  & 0.0141     & 0.0739  & 0.0403     \\
Skewness      & -0.1372 & 0.0829     & 0.1441  & -0.5957    \\
Kurtosis      & 2.4216  & 2.3507     & 1.7317  & 2.5336     \\
Jarque Bera   & 15.9656$^*$ & 17.4936$^*$    & 65.9032$^*$ & 63.7770$^*$ \\
\bottomrule
\multicolumn{5}{l}{$^*$ Significant at 1\% level.}\\
\end{tabular}
\end{table}

\begin{figure}[h!p]
    \centering
        \includegraphics[scale=0.55]{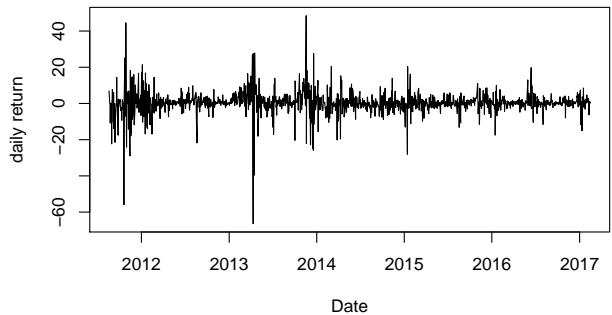}
    \caption{BTC daily return}
   \label{fig:dailyreturn}
\end{figure}

Our findings cover different aspects of these time series dynamics. First, the R/S method is biased to finding long memory in all the time series. It is unable to discriminate different dynamic regimes, present in the daily returns of the bitcoin market. On contrary, the DFA method clearly defines two periods: before and after 2014. In the first subperiod the time series of daily returns exhibits a persistent behavior, manifested in Hurst exponents greater than 0.5. In the second subperiod the Hurst exponents tend to wander around 0.5, making this behavior compatible with the EMH. According to \cite{Barivieraetal2017}, the causes for this regime switch are not yet found and returns' long memory is not related to trading volume. Second, the Hurst exponents corresponding to the series of return volatility exhibits long memory in all sliding windows. This kind of dynamics could produce the volatility clustering reflected in Figure \ref{fig:dailyreturn}. Third, the long term behavior of returns and volatility are different. This kind of behavior could hide some complex underlying dynamics, which exceeds the aim of this letter. Descriptive statistics of the Hurst exponents estimates are displayed in Table \ref{Hurst-statistics}.

\section{Conclusions \label{sec:conclusions}}
We study the long memory of the bitcoin market using the Hurst exponent, computed using two alternative methods. We advocate for the use of the DFA method because it is more robust and less sensitive to departures from stationarity conditions. We find that daily returns suffered a regime switch. From 2011 until 2014 the returns' time series was essentially persistent ($H>0.5$), whereas after that year, the behavior seems to be compatible with a white noise. On contrary, daily volatility (measured as the logarithmic difference between daily maximum and minimum price) exhibits a persistent behavior during all the period under study. In addition, the long memory content of daily volatility is stronger than in daily returns. This features give some hints on the characteristics of this synthetic currency market. In particular, that volatility clustering is a key feature of the Bitcoin market.

\bibliographystyle{apalike}
\bibliography{longmemorybib}   

\end{document}